
\magnification=\magstep1

\def\part{\partial}
\def\DD{\hbox{D}}
\def\wz{1} 
\def\dkw{2} 
\def\ml{3} 
\def\gp{4} 
\def\cl{5} 
\def\ll{6} 
\def\hans{7} 
\def\hp{8} 
\def\va{9} 
\def\no{10} 
\def\tink{11} 
\def\qcd{12} 
\def\bard{13} 
\def\ps{14} 
\def\fy{15} 
\def\pc{\phi_{\rm cl}}
\def\dd{\hbox{d}}
\def\de{\delta}
\def\ha{\hat a}

\def\he{\hat \eta}
\def\part{\partial}

\def\sql{\sqrt{\lambda}}
\footline={\hfil}
\font\ti=cmr7 scaled \magstep3
\rightline{IASSNS-HEP-92-18}
\rightline{Fermilab-92/126-T}
\rightline{hep-th@xxx/9205026}
\rightline{April 1992}
\vskip 1 truein
\centerline{{\ti Lorentz invariance of effective strings}}
\vskip 1 truein
\centerline{J.D. Cohn\footnote{${}^1$}{jdcohn@fnal.fnal.gov}}
\centerline{Fermilab, MS106}
\centerline{P.O. Box 500}
\centerline{Batavia, Illinois 60510}
\medskip
\centerline{and }
\medskip
\centerline{Vipul Periwal\footnote{${}^2$}{vipul@guinness.ias.edu}}
\centerline{The Institute for Advanced Study}
\centerline{Princeton, New Jersey 08540-4920}
\vfil
\par\noindent Abstract: Starting from a Poincar\'e invariant
field theory of a real scalar field with interactions governed by a
double-well potential in 2$+$1 dimensions,
the Lorentz representation induced on the
collective coordinates describing low-energy excitations
about an effective string background is derived.
In this representation, Lorentz transformations are given in terms of
an infinite series, in powers of derivatives along the worldsheet.
Transformations that act on the
direction transverse to the string worldsheet involve a universal
dimension $-1$ term.  As a consequence, Lorentz invariance holds in
this theory of long effective strings due to cancellations in the action
between
irrelevant terms and the dimension two term that describes
free massless scalar fields in two dimensions.

\vfil\eject
\footline={\hss\tenrm\folio\hss}
\centerline{1. Introduction}
\medskip
The equations of motion of several quantum field theories of physical
interest have classical static solutions that can be interpreted as
possessing string-like defects.  Such solutions spontaneously break
translation invariance in the $D-2$ transverse space dimensions.
As a result, there are Nambu-Goldstone
massless excitations about such backgrounds, even if the
field theory has only massive excitations about homogeneous
background configurations.  These massless modes have
wavefunctions supported in the vicinity of the
spacetime sheet swept out by the defect, so the effective field theory
describing low energy phenomena about such classical solutions is
two dimensional.  On general grounds, the leading
term in the action for these modes must take the form
$$S \approx S_o \equiv \hbox{const.}\ \int \dd t\dd y \left[
\left(\part_t f^i\right)^2
-\left(\part_y f^i\right)^2 \right],$$
where $S$ is the action of the underlying field theory, $S_o$
is the action governing low energy phenomena,
$i= 1, \dots , D-2,$ and we have chosen the $(t,y)$ plane
as the plane of the worldsheet.

The derivation of $S_o$ (and higher
terms) is standard, and excellent treatments are available
in the literature, including careful treatments of the introduction of
collective coordinates.  We have not, however, found a
calculation of the Lorentz transformations starting from the
underlying field theory.  In this note we do this
in a $2+1$ dimensional example and show that
Lorentz invariance in such theories
(of long effective strings)
is a consequence of cancellations between $S_o$ and terms that
are irrelevant for long-distance physics.

We consider the induced quantization of one string, neglecting
overhangs and many string (interacting or not) sectors.  Large overhangs
produce effects
suppressed by exponentials of the form $\exp(- \hbox{const.} mR),$
where $R$ is a length-scale characterizing the overhang.
For small overhangs ($R=O(1/m)$), the curvature ($\propto R^{-2}$)
becomes large enough that the interactions of the Goldstone bosons
cannot be neglected, or equivalently, the long distance
effective field theory description is not appropriate.

Section two is a brief review  of the effective action
governing $f.$  Section three provides a quick derivation of the
Lorentz algebra, section four concludes.

The problems afflicting massless
fields in two dimensions are not relevant to the discussion.
It will be assumed that we are just above the roughening
transition, or that appropriate boundary conditions are in place.

\bigskip
\centerline{2. Review\footnote{${}^*$}{See Wallace and Zia[\wz], %
Diehl, Kroll %
and Wagner[\dkw], and L\"uscher[\ml] for much of the material in this %
section.  }}
\medskip

The specific example is a real scalar field in 2$+$1
dimensions, with an action:
$$ S \equiv {1\over 2} \int \dd^3x \left[ \part_r\phi\part^r\phi
- \lambda (\phi^2-{m^2\over \lambda})^2\right].$$
The metric $\eta_{ij}$
has signature $+--,$ and the coordinates are $x^r\equiv(t,y,z)
\equiv(y^\mu,z).$
The equation of motion is
$$\part^2\phi +2\lambda(\phi^2 -{m^2\over \lambda})\phi = 0.$$
Static configurations invariant under translations in
the $y$ direction are solutions of
$$ - \part_z^2 \phi(z) + 2 \lambda \phi(z)^3 - 2 m^2 \phi(z) =0 .$$

\def\sech{\hbox{sech}}
The classical solution corresponding to one string (or domain wall) is
$$\pc(x) \equiv\ {m \over \sqrt \lambda}  \tanh mz.\eqno(1)$$
There are other solutions to the boundary conditions
$\phi \rightarrow \pm {m /\sql} (z\rightarrow\pm\infty),$ but these are
multi-string configurations.  
The spectrum of fluctuations about the kink solution is known and
corresponds to
eigenvalues of the operator $\Omega= -(\part_z-2\sql\pc)(\part_z+2\sql\pc),$
$$\eqalign{\lambda_0 = 0 &: \psi_0 =
{\sqrt{3m}\over 2} { }\sech^2(mz)
\equiv{\sqrt{3\lambda\over 4m^3} } \eta_0(z) ,\cr
\lambda_1 = 3m^2 &: \psi_1 = \sqrt{3m\over 2}  \sech(mz)\tanh(mz),\cr
\lambda_k = k^2+4m^2 &: \psi_k = {{\sqrt m\exp(ikz)}\over
{\sqrt{k^4+5k^2m^2+4m^2}}}
\left[3\tanh^2(mz)-{3ik\over m}\tanh(mz)-({k^2 \over m^2}+1)\right].\cr
= m_k^2 & \cr}$$
The physical significance of these
modes is as follows:
\item{(a)} The zero-mode is the Nambu-Goldstone boson,
since $\eta_0 =\part_z\pc.$
\item{(b)} The other localized mode has mass $\sqrt{3} m$ and is
referred to as the kink `excitation' in the literature.
It corresponds to a squeezing of the string.  To see this, compute
the normalized overlap of $z\dd\pc/\dd z$ and $\psi_1.$
This is $\pi\sqrt3/\sqrt{8\pi^2-48} \approx 0\cdot978.$
\item{(c)} A continuum starting at mass $2m,$ with $k$ taking
arbitrary real
values---these modes are the counterparts of the spectrum obtained when
expanding about  a homogeneous background $\pc(x) = \pm m/\sql.$

A single classical
solution with fixed kink position is not a good ground state to
quantize around because of the massless mode corresponding to moving
the wall.   Instead one introduces a
variable describing the position
of the wall and integrates over it. The position of the wall is a collective
coordinate.  It can be introduced in the path integral fomulation
derived by Gildener and Patrascioiu[\gp] of the implicit collective
coordinate method due to Christ and Lee[\cl].
Writing
$$\phi(t,y,z) \equiv \pc\left(z-f(t,y)\right) +
\xi\left(t,y,z-f(t,y)\right),$$
where
$$
\xi(t,y,z+\alpha) = a_0(t,y) \psi_0(z+ \alpha) +
a_1 (t,y) \psi_1(z+ \alpha) + \int dk a_k (t,y) \psi_k(z+\alpha)
$$
shows the coefficient of the zero mode explicitly.  The integral
over $k$ is schematic, it is not necessary to be precise since
loop effects will not be considered.  Inserting $\prod df(t,y)$
$\delta(g(f))|\de g/\de f| = 1 $ into the path integral, $g(f) =$
$\int dz \part_z \phi_{cl}(z-f) \phi(z)$,
the functional integral becomes:
$$\int\DD f\DD\xi \Delta[\xi] \prod_{t,y} \delta\left[
\int \dd z' \eta_0(z')\xi(z')\right] \ \hbox{e}^{iS}.$$
Here
$$\Delta\equiv \prod_{t,y}\int \dd z' \eta_0\left(z'-f(t,y)\right)
\part_{z'}\phi(t,y,z')= \Delta(a) $$
is actually independent of $f$ and $a_0$.
The functional integral for $\xi$ is defined, as usual, as
an integral over the coefficient
fields, $a_i.$  The integral over $a_0$ then
eliminates the delta function, and all dependence on $a_0$,
and the collective coordinate field, $f,$ is left in its stead.
Thus, $\xi$ can be treated henceforth as if $a_0=0.$
$\Delta(a)$ enters
the action at order $\hbar$ and so will be neglected in the following.
Terms in the perturbative expansion arising from $\Delta$ need
to be regulated.  Its field independent term is a
constant, so using dimensional regularization $\Delta$
can be set to one[\wz].

The action now takes the form
$$\eqalign{S = - {m\over \lambda}\int \dd^3x \bigg\{ \eta_0^2 -
{1\over 2} \part_\mu f\part^\mu f\left[\eta_0 + \part_z\xi\right]^2
&-{1\over 2} \xi\left(-\part_\mu\part^\mu +\part_z^2 -6\pc^2 +2 \right)\xi
\cr
&+2\pc\xi^3 + {1\over 2} \xi^4 + \part_\mu\xi\part^\mu
f\part_z\xi\bigg\} .\cr}$$
All the $f$ dependence is explicit and
we have rescaled fields and coordinates so that all dependence on
the parameters $m$ and $\lambda$ appears, as it must, in the
dimensionless combination $m/\lambda.$

\def\lraN#1#2#3{\langle #1|#2|#3\rangle}
\def\lran#1#2{\langle #1|#2\rangle}
\def\half{{1\over 2}}
\def\ho{\hat\Omega}

It is convenient in the
some of the following to work with the components of $\xi$ decomposed in
terms of the normalized wavefunctions, $\psi_i, i=1, \dots.$
Define $\he \equiv (\psi_1,\psi_k)$ and $\ha \equiv (a_1,a_k)$ as vectors,
so $\xi = \ha\cdot\he,$ and let $\ho$ be
the mass matrix, which in the $\psi$ basis, is
$\hbox{diag}(3,k^2+4)).$
Integrating out $z$, $S$ is now
$$\eqalign{-{m\over \lambda}\int \dd t\dd y \bigg\{\lran 00 &- \half
\part_\mu f\part^\mu f
\left[\lran 00 +2 a_k\lraN 0{\part_z}k -
a_ja_k\lraN j{\part_z^2}k \right]
+\half a_i\left( \part_{\mu}\part^{\mu} +\ho
\right)a_i \cr
&+2 \lraN i{\psi_j \phi_{cl}}k
a_ia_ja_k
+\half \lraN i{\psi_j \psi_k}l a_ia_ja_ka_l+
\part_\mu f\part^\mu
a_j\lraN j{\part_z}k a_k\bigg\}, \cr}$$
where $\lraN i{g(z)}j \equiv \int \dd z \psi_i(z) g(z) \psi_j(z)$.
Note that $|0\rangle $ will denote $\eta_0(z),$ which is {\it not}
normalized, $\int \eta_0^2(z) \dd z = 4/3.$  This exact rewriting of
the action in
the one string sector is a two dimensional field
theory with a
single massless field $f(t,y)$, the position of the wall, interacting
with an infinite number of massive fields $\{ a_i(t,y) \}$.

The equation of motion for $\ha$ is
$$
\eqalign{\big[-\part_\mu\part^\mu -\ho -(\part f)^2 \part_z^2 +2
\part^\mu f\part_\mu\part_z
&+ (\part_\mu\part^\mu f)\part_z\big]\xi \cr &= \part_z\eta_0(\part f)^2 +
6\pc\xi^2 + 4\xi^3,\cr}
$$
which can be solved, since $\ho$ is invertible,
to obtain $\ha$ as a series in $\part_\mu f,$ a useful expansion in the
long wavelength limit.
One obtains $\xi = \xi^{(2)} + \xi^{(4)} + \xi^{(2,2)} + \dots,$
where
$$\eqalign{\xi^{(2)} &= - \ho^{-1} \part_z\eta_0 (\part f)^2,\cr
\xi^{(4)} &= - \ho^{-1} \left[\part_z^2 (\part f)^2
+ 6\pc \xi^{(2)}\right]\xi^{(2)},\cr
\xi^{(2,2)} &= \ho^{-2}\part_z\eta_0 \part_\mu\part^\mu(\part f)^2.\cr}$$
In components, this amounts to
$$\eqalign{a_i =
&-\left\{\lraN i{\ho^{-1}\part_z}0 - \lraN i{\ho^{-2}\part_z}0 \part^2
\right\} (\part f)^2 \cr
&+\left\{\lraN i{\ho^{-1}\part_z^2\ho^{-1}\part_z}0 + 6
\lraN i{\psi_j\pc}k
\lraN j{\part_z}0
\lraN k{\part_z}0 \right\} (\part f)^4 + \dots \ .\cr}
$$


Using the identity of [\dkw],
$$
\part_z = {1 \over 2} [z, \Omega]\ \Leftrightarrow\
<i|\part_z|j> = {1 \over 2} (m_j^2 - m_i^2) <i|z|j>,
$$
and substituting for $\xi$, ({\it i.e.}, integrating
out $\xi$), we find (showing up to $O(\part^8)$)
$$\eqalign{
S  = - {m\over\lambda} \int \dd^2y \bigg\{&\lran 00\left[1- \half (\part f)^2
-{1\over 8}(\part f)^4 -{1\over 16} (\part f)^6 + \dots \right]\cr
+ &{1\over 8}\lraN 0{z^2}0 (\part f)^2\part_\mu\part^\mu (\part f)^2 +
\dots \bigg\}.\cr }\eqno(2)
$$
The first four terms, as shown by [\dkw],
give the leading terms in the expansion of $\sqrt{1 - (\part f)^2}$,
the Nambu-Goto action, with induced metric
$h_{ij} =\eta_{ij} - \part_i f \part_j f$.
The last term shown can be rewritten partly as
the curvature of the induced metric, but there are additional terms
as well, which do not appear to have a geometric interpretation.
Since the coefficient of this term is non-universal, the
appearance of non-geometric terms is not surprising---it does appear to
contradict the work of Ref.~\ll.  It may be that different
parametrizations of the collective coordinates lead to different
non-universal terms.

\bigskip
\centerline{3. Lorentz transformations}
\medskip
The canonical Lorentz generators are
$$ M_{rs} \equiv  \int \dd y\dd z \left[j_{0r}x_s - j_{0s}x_r\right].$$
where
$$j_{rs} \equiv -\eta_{rs} {\cal L} + \part_r\phi\part_s\phi$$
are the translation currents.

An arbitrary variation of $\phi$ can be written as
$$\eqalign{
\de \phi(z) =& -\de f \part_z \phi(z) + \de\ha \cdot \psi(z-f(t,y)) \cr
           =& \de a_0 \psi_0(z-f(t,y)) + \de \tilde{\ha}_i \psi_i(z-f(t,y)) \cr
}
$$
since a complete basis $(a_0, \tilde{\ha}_i)$
is dual to the complete set of $\psi_i$.
We then have
$$ \left(\matrix{\de a_0 \cr
    \de \tilde{\ha}_i\cr}\right) =
\left(\matrix{-\int dz \part_z \phi(z) \psi_0(z-f(t,y)) & 0 \cr
-\int dz \part_z \phi(z) \psi_i(z-f(t,y)) & 1 \cr}\right)
\left(\matrix{\de f \cr
    \de {\ha}_i\cr}\right) $$
Inverting this
gives $P_\phi\equiv-i\de/\de\phi,$ the momentum conjugate
to $\phi,$ in terms of
the momenta conjugate to $f$ and $a_i,$
$$
{\de \over \de \phi} =
{-\eta_0(z-f(t,y)) \over \Delta(\ha)} {\de \over \de f}
+
\left[-{{\eta_0(z-f(t,y))\ha_k\langle i|\part_z|k\rangle}
\over \Delta(\ha)} +
\psi_i(z-f(t,y)) \right] {\de \over \de \ha_i} \; .
$$
One can verify that [$P_{\phi}(t,y,z),\phi(t,y',z')$] =
$-i \de (z-z')\de (y-y')$ .

Ordering ambiguities do not change the commutation relations to leading
order in $\hbar$.  In the present context the
possible ordering ambiguities which are subleading are also not of concern
because it is possible
to regulate the theory without violating any relevant symmetry.

With these expressions at hand, it is straightforward to evaluate
the action of the Lorentz generators on $f$ and $a_i$:
$$
\eqalign{
[M_{0y},f]&= i (t \part_y f + y \part_0 f) \cr
[M_{0y},a_j]&= i (t \part_y a_j + y \part_0 a_j) \cr
[M_{\mu z},f]&= i \bigg[-y_\mu +f \part_\mu f +
{\part_\mu f \over \Delta(\ha)} a_j\langle 0|z \part_z|j\rangle -
{\part_\mu a_j \over \Delta(\ha)} \langle0|z|j\rangle\bigg] \cr
[M_{\mu z},a_j]&= i\bigg[\part_\mu a_k\langle j|z|k\rangle
- \part_\mu a_i a_k
{\langle 0|z |i\rangle\langle j|\part_z|k\rangle\over \Delta(\ha)} \cr
&-\part_\mu f \left( \langle j|z|0\rangle +
\langle j|z \part_z|k\rangle a_k
-a_i a_k
{{\langle 0|z \part_z|k\rangle\langle j|\part_z|i\rangle}\over
\Delta(\ha)}\right)
+ f \part_\mu a_j\bigg] \cr
}
$$
These are valid to leading order in $\hbar.$  One can substitute
$\xi=\xi(f),$ as derived in section two, to obtain the complete
nonlinear transformations that leave the effective action governing
$f$ (eq.~2) invariant.  This substitution only affects Lorentz
transformations of $f$ beginning at order $\part^3.$
The first two terms in the transformation of $f$ are independent of the
details of the wavefunctions and of the potential, and are thus universal
(if the kinetic term is canonical, for other possibilities see [\hans]).
They also leave the measure invariant, even though they are nonlinear.
The supersymmetric version of the universal part of the transformations
was found in [\hp], using the Volkov-Akulov formalism[\va].

\smallskip
\centerline{4. Concluding remarks}
\smallskip

The computations given above are entirely straightforward, and
nothing untoward or unexpected was found.
The result is
the complete, albeit intractable and impractical,
form of the Lorentz transformations, to all orders in the
derivative expansion, and leading order in $\hbar$.

This differs from the
Lorentz algebra of the fundamental string in light cone
gauge\footnote{${}^\dagger$}{For the particular case of $2+1$ %
dimensions, the light-cone Lorentz algebra is nonanomalous since it has only %
one nonlinear generator.  Inconsistencies appear at the level of %
interactions.}.  In that case, $S_o$ is the full action after solving
the constraints.  Upon quantizing, Lorentz transformations
require compensating conformal transformations to close
unless $d=2,3$.  These conformal transformations are not
symmetries in the quantized theory unless $d = 26$.
The coordinates used
in our example, $X^r = (t,y,z+f(t,y))$ do not obey the light cone
conditions $(\dot{X} \pm X')^2= 0$, the underlying field theory was
not quantized in light cone gauge.
A direct comparison of the algebra induced from the field theory with that of
fundamental strings involves the quantization of an
interacting scalar field theory in light-cone coordinates,
a difficult task
(aside from integrable theories in two dimensions,
which are in some sense free).

Here no conformal invariance is assumed or
used.\footnote{${}^\ddagger$}{While this work was %
being prepared for publication, A. Shapere pointed Ref.~[\hp] out to us where %
some of these observations are independently made, cited as %
J. Polchinski (unpublished).}
The variations of irrelevant terms cancel the
variation of $S_o,$ because the nontrivial Lorentz transformations of $f$
start with an inhomogenous universal term of dimension $-1.$  The light
cone gauge fundamental string can also be described by $S_o$, but
a fixed point theory describes the long (or short)
distance behaviour of an entire universality class
of theories---it does not follow that a global symmetry, {\it e.g.}
Lorentz invariance,
that appears in a given element of the universality class is common
to every other element.

To consistently
look at higher orders in $\hbar$ requires summing over loops in the underlying
field theory, finding a stringlike solution of the effective action
(rather than potential) and then repeating the procedure of eliminating the
massive fields in the derivative expansion.
(Repeating from earlier, there will also be
contributions from the collective coordinate Jacobian at higher orders,
which depend on the regulator chosen.  These contributions vanish in
dimensional regularization.)  All the terms
obtained by using the equations of motion to eliminate massive
excitations are explicitly local.  It is not clear to us that
the Polyakov--Liouville
term formed out of the induced metric can be made local in the gauge
choice inherited from the underlying field theory.

Classical solutions depend on the
parameters in the action, in the case above on $m$ and $ \lambda$.
Modifying these
parameters alters the width of the string and its string tension.
A question of interest is: are there values of these
parameters that lead to the decoupling of all fluctuations, other
than the Goldstone mode?  In other words, when does the quantum
theory, expanded about a classical solution with a string defect,
exhibit the characteristics of a structureless fundamental string?
Nielsen and Olesen[\no] addressed this question
and argued that the string is effectively of zero width
when the length scale for the energy levels for excitations of the string
(set by the string tension $\alpha '$) is much greater than the
length scales characterizing the width of the string (the penetration
depth $\lambda$ and the correlation length $\xi$ in the Abelian Higgs model),
{\it i.e.} $\sqrt{\alpha '} \gg \lambda , \xi$.  In the Abelian Higgs
model, this requirement translated into the electric charge,
$e \gg 1$.  (For flux tubes with more flux the effective charge
decreases.)  This implies that the physical paradigm
for their model, flux tubes in strongly type II superconductors,
are not well described by structureless fundamental
strings either[\tink].
Chromoelectric flux tubes in
lattice QCD have $\sqrt{\alpha'}$ of about the same order of magnitude
as the transverse width, $0\cdot 5$ fm[\qcd].

The example discussed in this paper can be studied with the same
criterion, putting back in the original
$\lambda, m$ dependence
of the fields and coordinates.
The classical string solution
in equation (1)
has a characteristic width $m^{-1}$, of the order
of the mass of fluctuations about homogeneous backgrounds.
The other scale is
the energy density of the classical solution which sets the
energy scale for excitations of the string:
$$
{1 \over \alpha '} \propto
\int \dd z \left(\part_z \phi_{cl}(z)\right)^2 \propto
{m^3 \over \lambda} \; \; .
$$
The string is `thin' when fluctuations of the string will
not excite the internal modes, {\it i.e.}
$$
m \gg{1 \over \sqrt{\alpha '}} \propto {m^{3/2} \over \sql}
\ \Leftrightarrow \ \lambda \gg m \; .
$$
So in this case as well, the thin string condition
is the strong coupling limit.  As Nielsen and Olesen pointed out,
this is the $\hbar \rightarrow \infty$ limit,
which makes `classical field considerations very doubtful'.  For
the example studied here, this limit
(in Euclidean space) is in
the universality class of the high temperature ferromagnetic Ising model,
where entropic considerations dominate over
energetic considerations.  Thus although an isolated sector of the theory
with a single string is an energetically preferred classical configuration,
the fluctuations around it are so large (`$1/\hbar$' is small) that
it is not a good approximation to the most probable state
of the system\footnote{${}^*$}{In Ref.~[\dkw] %
it is argued that one can
neglect higher loops in the strong coupling limit.  This seems to %
be incorrect; no details are given, however, so this may be due to %
misunderstanding on our part.}.

The above discussion gives little insight into the
quantization of the Nambu-Goto area action in dimensions other than
26.  There are various suggestions in the literature, such as
including the effects of `kinks'[\bard] (making the string massive,
quantizing, and then taking the massless limit), or
adding to $S_o$ a non-polynomial term in the $f^i$'s,
and then using conformal invariance to fix its coefficient[\ps].
For a further study of the proposal in [\ps],
see [\fy].
As was first argued by Nielsen and
Olesen[\no], and also seen in the example studied here,
the thin string limit
(where an area interpretation of the action might apply)
corresponds to strong coupling.  In this limit the Lorentz
transformations here are renormalized, but beyond this it is hard
to make definitive statements.
Although problems in quantizing the
Nambu-Goto string outside of the critical dimension
may manifest themselves as failures of
Lorentz invariance, there is no inconsistency in the
Lorentz transformation properties of effective strings in any
regime where their existence can be reliably assumed.

\vfill\eject
Acknowledgments: We are grateful to A. Shapere for telling us about
Ref.~\hp, to M. Ro\v cek for discussions,
to J. Polchinski for e-mail communications, and
especially A. Strominger for many helpful comments.
J.D.C. also thanks A. Albrecht, L. Brekke, M. Douglas,
G. Harris, J. Lykken, E. Raiten, D. Spector, S.-H. Tye and N. Weiss
for discussions and
V.P. thanks D. Gross and E. Witten for their interest.
V.P. was supported by D.O.E. grant $\#$DE-FG02-90-ER40542.
\bigskip
\centerline{References}
\medskip
\item{\wz.} D.J. Wallace and R.K.P. Zia, {\sl Phys. Rev. Lett.} {\bf 43}
(1979) 808
\item{\dkw.} H.W. Diehl, D.M. Kroll and H. Wagner, {\sl Z. Physik B}
{\bf 36} (1980) 329
\item{\ml.} M. L\"uscher, {\sl Nucl. Phys.} {\bf B180} (1981) 317
\item{\gp.} E. Gildener and A. Patrascioiu, {\sl Phys. Rev.} {\bf D16}
(1977) 423
\item{\cl.} N.H. Christ and T.D. Lee, {\sl Phys. Rev.} {\bf D12} (1975)
1606
\item{\ll.} S.C. Lin and M.J. Lowe, {\sl J. Phys. A} {\bf 16} (1983) 347
\item{\hans.} T. Hansson, J. Isberg, U. Lindstr\"om, H. Nordstr\"om, J.
Grundberg, {\sl Phys. Lett.} {\bf 261B} (1991) 379
\item{\hp.} J. Hughes and J. Polchinski,
{\sl Nucl. Phys.} {\bf B278} (1986) 147
\item{\va.} D.V. Volkov and V.P. Akulov, {\sl JETP Lett.} {\bf 16} (1972) 438
\item{\no.} H.B. Nielsen and P. Olesen, {\sl Nucl. Phys.} {\bf B61} (1973) 45
\item{\tink.} For a calculation of the string tension in a strongly type
II superconductor, see for example
M. Tinkham, {\it Introduction to Superconductivity},
(Kreiger:1980, USA), p.~148.
\item{\qcd.} See for example A. DiGiacomo, M. Maggiore, S. Olejnik,
{\sl Nucl. Phys.} {\bf B347} (1990)441
\item{\bard.} W. Bardeen, I. Bars, A.J. Hanson, R.D. Peccei,
{\sl Phys. Rev.} {\bf D13} (1976) 2364; {\bf D14} (1976) 2193.
\item{\ps.} J. Polchinski and A. Strominger, {\sl Phys. Rev. Lett.} {\bf
67} (1991) 1681
\item{\fy.} P.H. Frampton and M. Yamaguchi, Univ. of North Carolina
preprint IFP-420-UNC (1992)

\end